\documentclass[12pt,reqno]{amsart}  

\usepackage{a4}
\usepackage{amsmath}
\usepackage{amssymb}
\usepackage{epsfig}                 
\newcommand{\color}[2][{}]{}        

\theoremstyle{plain}            
\newtheorem{theorem}{Theorem}[section]
\newtheorem*{theorem*}{Theorem}

\newtheorem{lemma}[theorem]{Lemma}

\theoremstyle{definition}       

\theoremstyle{remark}
\newtheorem{remark}[theorem]{Remark}

\numberwithin{equation}{section}

\DeclareMathOperator{\dom}    {dom}

\DeclareMathOperator{\supp}   {supp}

\newlength{\maxbreite}%
\newlength{\maxhoehe}%
\newlength{\maxtiefe}%

\newcommand{\stelldrueber}[3][0pt]{
  \settowidth{\maxbreite}{#3}%
  \settoheight{\maxhoehe}{#3}%
  \settodepth{\maxtiefe}{#2}%
  \addtolength{\maxhoehe}{\maxtiefe}%
  {\makebox[\maxbreite]{\raisebox{\maxhoehe}{\hspace{#1}#2}}%
  \makebox[0pt][r]{#3}}%
}

\newcommand{\overcirc}[1]       
{\stelldrueber[.45ex]{$\scriptscriptstyle\circ$}{${#1}$}}

\newcommand{\R}{\mathbb{R}} 
\newcommand{\Sphere}{\mathbb{S}} 
\newcommand{\eps}{\varepsilon} 
\renewcommand{\phi}{\varphi}   

\newcommand{\HS}{\mathcal H}                

\newcommand{\Cci}[1]{C_{\mathrm c}^\infty ({#1})} 
\newcommand{\Lsqr}[1]{L_2({#1})}            
\newcommand{\lsqr}[1]{\ell_2({#1})}            
\newcommand{\Sob}[2][1]{\HS^{#1}({#2})}     
\newcommand{\Sobn}[2][1]{{\overcirc {\mathcal H}{}^{#1}({#2})}}
\newcommand{\norm}[2][{}]{\|{#2}\|_{#1}}    
\newcommand{\normsqr}[2][{}]{\|{#2}\|^2_{#1}} 
\newcommand{\iprod}[3][{}]{\langle{#2},{#3}\rangle_{#1}}  
\newcommand{\bd}  {\partial}                

\newcommand{\restr}[1]{{\restriction}_{#1}} 
\newcommand{\orth}{\bot}                    
\newcommand{\map}[3]{ #1 \colon #2 \longrightarrow #3} 
\newcommand{\set}[2]{\{ \, #1 \, | \, #2 \, \} } 

\usepackage{bbm}         

\newcommand{\Neu}{{\mathrm N}}              
\newcommand{\Dir}{{\mathrm D}}              

\newcommand{\laplacianD}[1]{\Delta^\Dir_{{#1}}}
\newcommand{\laplacianDN}[1]{\Delta^{\Dir \Neu} _{{#1}}}
\newcommand{\EW}[2]{\lambda_{#1}({#2})}     

\newcommand{\EWD}[2]{\lambda^\Dir_{#1}({#2})}
\newcommand{\EWDN}[2]{\lambda^{\Dir \Neu}_{#1}({#2})}

\newcommand{\Mnull}{{M_0}}                    
\newcommand{\Meps}{{M_\eps}}                  

\newcommand{\geps}{{g_\eps}}                  

\newcommand{\Ij}{{I_j}}                       

\newcommand{\Ueps}{{U_\eps}}                  
\newcommand{\Uepsj}{{U_{\eps, j}}}            
\newcommand{\Uj}{{U_j}}                       
\newcommand{\tUeps}{{\tilde U_\eps}}          

\newcommand{\Vepsk}{{V_{\eps, k}}}            
\newcommand{\Vk}{{V_k}}                       

\newcommand{\Aepsj}{A_{\eps, j}}            

\newcommand{\Feps}{F_\eps}

\begin{document}

\title[Branched quantum wave guides]
      {Branched quantum wave guides with Dirichlet boundary conditions: 
         the decoupling case}
      
      \date{\today}

\author{Olaf Post}      
\address{Institut f\"ur Reine und Angewandte Mathematik,
       Rheinisch-Westf\"alische Technische Hochschule Aachen,
       Templergraben 55,
       52062 Aachen,
       Germany}
\email{post@iram.rwth-aachen.de}
\date{\today}




\begin{abstract}
  We consider a family of open sets $\Meps$ which shrinks with respect
  to an appropriate parameter $\eps$ to a graph. Under the additional
  assumption that the vertex neighbourhoods are small we show that the
  appropriately shifted Dirichlet spectrum of $\Meps$ converges to 
  the spectrum of the (differential) Laplacian on the graph with
  Dirichlet boundary conditions at the vertices, i.e., a graph
  operator \emph{without} coupling between different edges. The
  smallness is expressed by a lower bound on the first eigenvalue of a
  mixed eigenvalue problem on the vertex neighbourhood. The lower
  bound is given by the first transversal mode of the edge
  neighbourhood.  We also allow curved edges and show that all bounded
  eigenvalues converge to the spectrum of a Laplacian acting on the
  edge with an additional potential coming from the curvature.
\end{abstract}

\maketitle



\section{Introduction}

Graph models of quantum systems can often be used to describe in a
simple way some important aspects of the behaviour of a quantum
system. Although such models are simple enough to be solvable (because
they are essentially $1$-dimensional) they still have enough structure
to model real systems.  Ruedenberg and Scherr
\cite{ruedenberg-scherr:53} used this idea to calculate spectra of
aromatic carbohydrate molecules.  Nowadays the rapid technical
progress allows to fabricate structures of electronic devices where
quantum effects play a dominant role.  Graph models like quantum
graphs (also called \emph{metric graphs}) can often be viewed as a
good approximation of such structures. From the mathematical point of
view these models were analysed first thoroughly in
\cite{exner-seba:89} , for recent developments, bibliography and
further applications see \cite{duclos-exner:95},
\cite{kostrykin-schrader:99},~\cite{kuchment:02}
or~\cite{kuchment:04}; note that \cite{karp-pinsky:88} also calculated
the eigenvalue asymptotic of a tubular $\eps$-neighbourhood of a
curve.

A quantum (or metric) graph is a graph where we associate a length to
each edge. A natural operator acting on such graphs is given by a
self-adjoint extension of $-d^2/dx^2$ on each edge. We will call such
a self-adjoint extension a \emph{Laplacian} on the (quantum) graph.
Note that the Laplacian on a \emph{discrete} graph is a
\emph{difference} operator on $\lsqr K$ rather than a
\emph{differential} operator acting in $\bigoplus_j \Lsqr {e_j}$.
Here, $K$ labels the vertices and $J$ the edges $e_j$, $j \in J$, of
the graph. A detailed overview on this wide field can be found in
\cite{kuchment:04} or \cite{kostrykin-schrader:99}.

A natural question is in what mathematical sense a quantum graph $M_0$
can be approximated by a more smooth space $\Meps$. One is interested
what Laplacians on $M_0$ occur as limit operators from operators on
$\Meps$. More significantly, $\Meps$ could be the $\eps$-neighbourhood
of an embedded graph $M_0 \subset \R^n$ or a manifold shrinking to
$M_0$ as $\eps \to 0$. We call such approximating spaces
\emph{branched quantum wave guides}.  Recently, spectral convergence
in the case of a bounded open set $\Meps$ with Neumann boundary
condition has been established
in~\cite{rubinstein-schatzman:01},~\cite{kuchment-zeng:01}
and~\cite{kuchment-zeng:03}; for an approximation by manifolds
see~\cite{exner-post:pre03}. All these examples have in common, that
the lowest eigenmode of the transversal direction is $0$ with constant
eigenfunction. In this case, the limit operator is the Laplacian on
the graph with \emph{Kirchhoff boundary conditions}, i.e., a function
$f$ in the domain of the Kirchhoff Laplacian is continuous at each
vertex and satisfies
\begin{equation}
  \label{eq:kirchhoff}
  \sum_{j \in J_k} f_j'(v_k) = 0, \qquad k \in K.
\end{equation}
In addition, the spectral convergence holds \emph{independently} of a
given embedding of the graph.  In particular, the convergence is
independent of the curvature of the embedded edges.

The case of an approximation by Dirichlet Laplacians on an open set
$\Meps$ was first treated heuristically
in~\cite{ruedenberg-scherr:53}. This case is harder to analyse since
the first transversal eigenvalue equals $\EWD 1 \Feps =
\lambda_1/\eps^2$ ($\lambda_1 > 0$), i.e., it is of the order
$\eps^{-2}$ if $\eps$ denotes the radius of the cross section
$\Feps=(-\eps,\eps)$ of the approximating set $\Meps$. A rescaling is
necessary, and first order terms of the metric $g_\eps$
(cf.~\eqref{eq:met.curv}) like the curvature become important. In
particular, the curvature of the (embedded) edge enters in the limit
operator as an additional potential.

\subsubsection*{Main result}

\begin{figure}[h]
  \begin{center}
\begin{picture}(0,0)%
\includegraphics{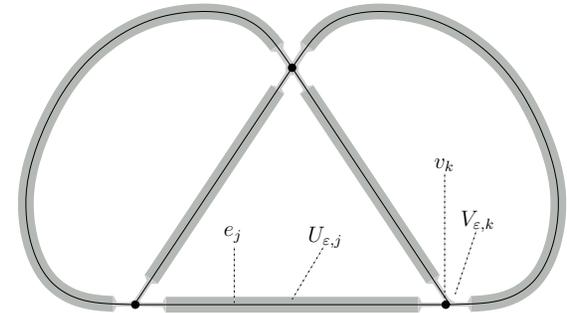}%
\end{picture}%
\setlength{\unitlength}{4144sp}%
 \begin{picture}(4768,2794)(317,-2042)
 \put(4141,-1186){$\Vepsk$}
 \put(2836,-1321){$\Uepsj$}%
  \put(3916,-691){$v_k$}
  \put(2116,-1276){$e_j$}
 \end{picture}%
    \caption{Decomposition of the graph neighbourhood $\Meps$ (grey)
      of the graph $M_0$ into edge
      and vertex neighbourhoods $\Uepsj$ and $\Vepsk$.}
    \label{fig:nbhd}
  \end{center}
\end{figure}

Assume that $M_0 \subset \R^2$ is a finite graph.  Our aim in this
note is to show the spectral convergence of the \emph{Dirichlet}
Laplacian on an approximating open set $\Meps \supset M_0$.  We suppose
that $\Meps$ can be decomposed into neighbourhoods $\Uepsj$ of the
edges $e_j$ and into neighbourhoods $\Vepsk$ of the vertices $v_k$ of
$M_0$ (cf.~Figure~\ref{fig:nbhd}). We assume that $\Vepsk$ is
$\eps$-homothetic to a fixed set $\Vk$.
The precise definition will be given in Section~\ref{sec:prelim}
and~\ref{sec:curvature}. Our basic assumption is that the vertex
neighbourhoods $\Vepsk$ are \emph{small},
\begin{figure}[h]
  \begin{center}
\begin{picture}(0,0)%
\includegraphics{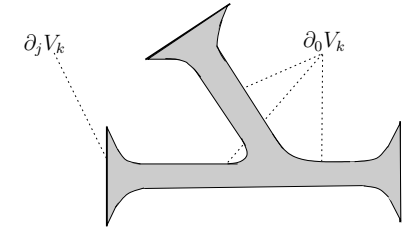}%
\end{picture}%
\setlength{\unitlength}{4144sp}%
\begin{picture}(3240,1946)(150,-1313)
\put(150,220){$\bd_j \Vk$}
\put(2540,240){$\bd_0 \Vk$}
\end{picture}%
    \caption{The scaled vertex neighbourhood $\Vk$ with the boundary part
       $\bd_0 \Vk$ coming from the original boundary and the boundary part
       $\bd_j \Vk$ where the edge $e_j$ emanates.}
    \label{fig:vertex}
  \end{center}
\end{figure}
i.e., that
\begin{equation}
  \label{eq:small}
  \EWDN 1 \Vk > \EWD 1 F = \lambda_1
\end{equation}
where $\EWDN 1 \Vk$ is the lowest eigenvalue of the Laplacian
$\laplacianDN \Vk$ of $\Vk$ with Dirichlet boundary conditions on
$\bd_0 \Vk$ (i.e., on the boundary induced from the original boundary
of $\Meps$) and Neumann boundary conditions on $\bd_j \Vk$, $j \in
J_k$, (i.e., on the parts where the adjacent edge neighbourhoods
labeled by $j \in J_k$ emanate, cf.~Figure~\ref{fig:vertex}).
Furthermore, $F=(-1,1)$ and therefore $\lambda_1=\pi^2/4$. We comment
on this condition in Section~\ref{sec:examples}.

Our main result is
\begin{theorem}
  \label{thm:main}
  Suppose that $\Meps$ is an open neighbourhood of a finite graph $M_0
  \subset \R^2$ satisfying the smallness assumption~\eqref{eq:small}
  on each vertex neighbourhood. Denote by $\EW k \eps$ the $k$-th
  eigenvalue of the Dirichlet-Laplacian $\laplacianD \Meps \ge 0$
  (counted with respect to multiplicity). Then
  \begin{equation}
    \label{eq:sp.conv}
    \EW k \eps - \frac {\lambda_1} {\eps^2} \to \EW k 0, \qquad \eps \to 0
  \end{equation}
  where $\EW k 0$ denotes the $k$-th eigenvalue of
  \begin{displaymath}
     \bigoplus_{j \in J} (\laplacianD {I_j} - \kappa_j^2/4)
  \end{displaymath}
  with $\laplacianD {e_j} = -d^2/dx_j^2$ being the Dirichlet Laplacian
  on the edge $e_j \cong I_j = (0,\ell_j)$ and $\kappa_j$ being the
  curvature of the embedded edge $e_j \subset \R^2$
  (cf.~\eqref{eq:curv}).
\end{theorem}

Note that the smallness assumption at the junctions $\Vepsk$ implies
that the limit operator \emph{decouples}, i.e., the limit operator is
the direct sum of operators acting on a single edge.  In the case of
the Neumann Laplacian on $\Meps$ decoupling occurs if the area of the
edge neighbourhood decays \emph{faster} than the area of the vertex
neighbourhood; e.g., if the latter scales in each direction of the
order $\eps^\alpha$ with $0 \le \alpha < 1/2$; the vertex
neighbourhoods are large obstacles seen from the edge neighbourhoods
(cf.~\cite{kuchment-zeng:03} or \cite{exner-post:pre03}).  In the case
of Dirichlet boundary conditions, in contrast, decoupling already
occurs when the vertex neighbourhoods scale with $\eps$, i.e., even
when the edge neighbourhood volume (which is of order $\eps$) decays
\emph{slower} than the vertex neighbourhood volume (of order
$\eps^2$).

We also show in Section~\ref{sec:examples} that the usual
$\eps$-neighbourhood $\Meps := \set{ z \in \R^2} {d(z,M_0) < \eps}$
does not satisfy our hypothesis since the leading order of the lowest
eigenvalue is at most $\mu/\eps^2$ with $\mu < \lambda_1$.  Therefore,
$\laplacianD \Meps - \lambda_1/\eps^2$ has a negative eigenvalue
tending to $-\infty$ (cf.~also Lemma~\ref{lem:pos}) and the conclusion
of Theorem~\ref{thm:main} fails. In particular, there is no limit
operator on the graph (using the simple shift $\lambda \to \lambda -
\lambda_1/\eps^2$), and the suggestion in~\cite{ruedenberg-scherr:53},
that the limit operator is the Kirchhoff Laplacian on $M_0$, is false
(cf.~also~\cite[Sec.~2.1 and~3.2]{kuchment:02}). Note that Ruedenberg
and Scherr implicitly assumed that the lowest eigenfunction does not
concentrate around the vertex which is the case as we will see in the
last section.

The spectral convergence of a single curved quantum wave guide has
already been shown in~\cite{duclos-exner:95} and~\cite{karp-pinsky:88}
using perturbation methods.  Our proof only uses variational methods
and is a simple adaption of~\cite{exner-post:pre03},
\cite{kuchment-zeng:01, kuchment-zeng:03} or
\cite{rubinstein-schatzman:01}, where one compares Rayleigh quotients.

The paper is structured as follows: In the next section we define the
required spaces and operators in the case of \emph{straight} edges.
Section~\ref{sec:graph} is devoted to the proof of
Theorem~\ref{thm:main} in this case. In Section~\ref{sec:curvature} we
provide the necessary changes in order to prove the result with
\emph{curved} edges.  Section~\ref{sec:examples} contains some
explantation on the smallness condition~\eqref{eq:small}, and examples
where this condition holds or fails.

\section{Preliminaries}
\label{sec:prelim}

In this section we define the limit space and the approximating space
together with the associated operators. We first consider
\emph{straight} edges without curvature, i.e., $\kappa_j=0$. In
Section~\ref{sec:curvature} we also allow curved edges.

\subsubsection*{Definition of the limit space}
Let $M_0$ be a finite connected graph with (metric) edges $e_j$, $j
\in J$ ($e_j$ isometric to an open interval $I_j=(0,\ell_j)$) and
vertices $v_k$, $k \in K$.  We denote the set of all $j \in J$ such
that $e_j$ emanates from the vertex $v_k$ by $J_k$.  The Hilbert space
associated to such a graph is
\begin{displaymath}
  \HS := \Lsqr \Mnull = 
  \bigoplus_{j \in J} \Lsqr \Ij
\end{displaymath}
which consists of all functions $f$ with finite norm
\begin{displaymath}
  \normsqr[0] f = \normsqr[\Mnull] f = \sum_{j \in J} \normsqr[\Ij] {f_j} =
  \sum_{j \in J} \int_{\Ij} |f_j(x)|^2 \,dx.
\end{displaymath}

\subsubsection*{Definition of the limit operator}
We define the limit operator $Q_0$ via the quadratic form
\begin{displaymath}
  q_0(f) := 
  \sum_{j \in J} \normsqr[\Ij] {f'_j} =
  \sum_{j \in J} \int_{\Ij} |f'_j(x)|^2 dx
\end{displaymath}
for functions $f \in \Cci \Mnull = \bigoplus_j \Cci \Ij$ (with compact
support).  The form closure of $q_0$ (also denoted by $q_0$) is the
extension of $q_0$ to the closure of the space of all such functions
in the norm
\begin{displaymath}
  \normsqr[0,1] f := \normsqr[0] f + q_0(f)
\end{displaymath}
(see \cite[Chapter~VI]{kato:66}, \cite{reed-simon-1} or
\cite{davies:96} for details on quadratic forms).  Note that
\begin{displaymath}
  \dom q_0 =  \bigoplus_{j \in J} \Sobn \Ij.
\end{displaymath}
Remember that $\Sobn I$ is the closure of $\Cci I$ w.r.t.~the norm
$(\normsqr f + \normsqr{f'})^{1/2}$. The associated self-adjoint,
non-negative operator $Q_0$ is given by
\begin{equation}
\label{eq:formaledge}
  Q_0  = \bigoplus_{j \in J} \laplacianD \Ij 
\end{equation}
where $\laplacianD \Ij \ge 0$ denotes the self-adjoint operator
$-d^2/dx^2$ on $\Ij$ with Dirichlet boundary conditions.  The spectrum
of $Q_0$ is purely discrete and will be denoted by $\EW k 0$, written
in ascending order and repeated according to multiplicity.

\subsubsection*{Definition of the approximating space}
We now describe the family of open sets $(\Meps)_\eps$,
$0<\eps\le\eps_0$, approximating the graph $M_0$ as $\eps \to 0$.  For
convenience only, suppose that $M_0$ is embedded in $\R^2$ (an
abstract definition of $\Meps$ in the general case will be given
soon). Assume that we can decompose $\Meps$ into open sets $\Uepsj$
containing those points $x \in e_j$ with $d(x,\bd e_j) > a_j \eps/2$ for
some real number $a_j < 1/\eps_0$ and $\Vepsk \ni v_k$ such that the
union of their closures equals $\overline \Meps$. Here, the \emph{edge
  neighbourhood} $\Uepsj$ is isometric to $I_{\eps,j} \times \Feps$
(both equipped with the Euclidean metric) where $I_{\eps,j} :=
(0,(1-a_j\eps)\ell_j)$ and $\Feps=(-\eps,\eps)$ is the scaled cross
section. Furthermore, we assume that the \emph{vertex neighbourhood}
$\Vepsk$ is $\eps$-homothetic to a fixed open set $\Vk$.  Using a
simple coordinate transform we have therefore the isometries
\begin{equation}
  \label{eq:isometries}
  (\Uepsj, g_\mathrm{eucl}) \cong (\Ij \times F, \geps) 
               \qquad \text{and} \qquad
  (\Vepsk, g_\mathrm{eucl}) \cong (\Vk, \geps)
\end{equation}
where
\begin{equation}
  \label{eq:metric}
  \geps = (1-\eps a_j)^2 dx^2 + \eps^2 dy^2 \qquad \text{and} \qquad
  \geps = \eps^2 g
\end{equation}
are the metrics on the edge resp.\ vertex neighbourhood. Here,
$F=(-1,1)$ and $g$ is the Euclidean metric on $\Vk$.  In the sequel we
use this change of coordinate transform without mentioning. Note that
the slightly shortened edge neighbourhood is necessary in order to have
an embedding for the edge \emph{and} the vertex neighbourhood.

Although we are mainly interested in the embedded situation as
described above, we prefer the following abstract setting in order to
keep the notation of~\cite{exner-post:pre03} and recognise the
important geometric objects (not depending on any embedding). For each
$j \in J$ we let $\Uepsj$ be the Riemannian manifold $(\Ij \times F,
\geps)$ where $\geps$ is given as in~\eqref{eq:metric}. Here $F$ is
the interior of a compact, connected $m$-dimensional manifold ($m \ge
1$) with metric denoted by $dy^2$ having purely discrete Dirichlet
spectrum with first eigenvalue $\lambda_1>0$.

Furthermore, we denote by $\Vepsk$ the Riemannian manifold $(\Vk,
\geps)$ with $\geps=\eps^2 g$ where $g$ is a metric on $\Vk$.  We
assume that
\begin{equation}
  \label{eq:bd.vk}
  \bd \Vk = \bd_0 \Vk \cup \bigcup_{j \in J_k} \bd_j \Vk,
\end{equation}
i.e., the boundary of $\Vk$ has as many boundary parts $\bd_j \Vk$
isometric to $F$ as edges emanate from $v_k$ and $\bd_0 \Vk$ is the
closure of $\bd \Vk \setminus \bigcup_j \bd_j \Vk$
(cf.~Figure~\ref{fig:vertex}). Furthermore, we assume that the metric
on $\Vk$ has product structure $g=dx^2 + dy^2$ near $\bd_j \Vk$.

We can define an abstract manifold $\Meps$ by identifying the
appropriate boundary parts according to the graph $M_0$. Note that a
smooth structure on $\Meps$ and also a smooth metric $g_\eps$ of the
form~\eqref{eq:metric} in the respective charts exist since $\Meps$ is
diffeomorphic to a product $(0,1) \times F$ in a neighbourhood of each
$\bd_j \Vk$ on both sides of $\bd_j \Vk$, i.e., on $\Vk$ and $\Uj$.
Strictly speaking we should introduce another chart for each $j \in
J_k$ and $k \in K$ covering $\bd_j \Vk$ in order to define the smooth
structure properly. But since we only use \emph{integrals} over
$\Meps$, a cover up to sets of measure $0$ is enough.  The resulting
manifold has dimension $d=m+1$. Note that $\Meps$ need not to be
embedded in any space, but the embedded case described above is also
covered by this setting.

The associated Hilbert space is
\begin{displaymath}
  \Lsqr \Meps = 
  \bigoplus_{j \in J} \Lsqr \Uepsj \oplus
      \bigoplus_{k \in K} \Lsqr \Vepsk
\end{displaymath}
which consists of all functions $u$ with finite norm
\begin{multline*}
  \normsqr[\eps] u =
  \normsqr[\Meps] u = 
  \sum_{j \in J} \normsqr[\Uepsj] u +
     \sum_{k \in K} \normsqr[\Vepsk] u \\ =
  \sum_{j \in J} \int_{\Ij \times F} |u|^2 (1-a_j\eps)\eps^m\, dx \,dy
     + \sum_{k \in K} \int_{\Vk} |u|^2 \eps^d dz
\end{multline*}
where $dy$ and $dz$ represent the natural measures on $F$ and $\Vk$,
respectively.

\subsubsection*{Definition of the operator on the manifold}
The operator on the thickened space we are considering will be the
Dirichlet Laplacian on $\Meps$, i.e., $H_\eps = \laplacianD \Meps \ge
0$.  The corresponding quadratic form $h_\eps$ is given by
\begin{multline*}
  h_\eps (u) = 
  \sum_{j \in J} \normsqr[\Uepsj] {d u} +
     \sum_{k \in K} \normsqr[\Vepsk] {d u} \\ =
  \sum_{j \in J} \int_{\Ij \times F}
     \Bigl[
         \frac 1 {(1-a_j\eps)^2} |\partial_x u|^2 +
         \frac 1 {\eps^2} |d_y u|^2 
     \Bigr]
        (1-a_j\eps)\eps^m\, dx \, dy
     + \sum_{k \in K}
           \int_{\Vk} |d u|^2 \eps^{d-2} dz
\end{multline*}
for functions $u \in \dom h_\eps = \Sobn \Meps$ where $\Sobn \Meps$ is
the closure of $\Cci \Meps$ in the norm $(\normsqr u +
h_\eps(u))^{1/2}$. Here, $|d_yu|^2$ and $|du|^2$ are evaluated in the
($\eps$-independent) metric of the exterior derivative of $u(x,\cdot)$
and $u$ on $T^*F$ and $T^*\Vk$, respectively.

The spectrum of $H_\eps$ is again purely discrete (since $\overline
\Meps$ is compact) and will be denoted by $\EW k \eps$, written in
ascending order and repeated according to multiplicity.  By the the
\emph{min-max principle} we have
\begin{equation}
  \label{eq:min.max}
  \EW k \eps =
  \inf_{L_k} \sup_{u \in L_k \setminus \{0\} }
      \frac {h_\eps(u)}{\normsqr[\eps] u},
\end{equation}
\sloppy where the infimum is taken over all $k$-dimensional subspaces
$L_k$ of $\dom h_\eps$, cf.~e.g.~\cite{davies:96}.

We denote by $\Feps$ the manifold $F$ with metric $\eps^2 dy^2$ and
the first Dirichlet eigenvalue of $F$ by $\lambda_1=\EWD 1 F>0$.
Since the lowest eigenvalue of $\Feps$ is $\lambda_1/\eps^2$, we need
a rescaling of the operator $H_\eps$ in order to expect convergence to
an $\eps$-independent limit operator. Therefore we set
\begin{equation}
  \label{eq:rescale}
  Q_\eps := H_\eps - \frac {\lambda_1}{\eps^2}
\end{equation}
and denote by $q_\eps$ the associated quadratic form.

We first note that the operator $Q_\eps$ is positive:
\begin{lemma}
  \label{lem:pos}
  Suppose the smallness condition~\eqref{eq:small} is fulfilled, then
  $Q_\eps \ge 0$.
\end{lemma}
\begin{proof}
  For $u \in \dom q_\eps=\dom h_\eps$ we have
  \begin{multline*}
    q_\eps(u) =
   \sum_{j \in J} \int_{\Ij}
     \Bigl[
         \frac 1 {(1-a_j\eps)^2} \normsqr[F] {\partial_x u(x,\cdot)} \\ +
         \frac 1 {\eps^2}
             \bigl(
                  \normsqr[F]{d_y u(x,\cdot)} - 
                  \lambda_1 \normsqr[F] {u(x,\cdot)}
             \bigr)
     \Bigr]
        (1-a_j\eps)\eps^m\,  dx\\ 
     +  \eps^{d-2} \sum_{k \in K}
     \Bigl[
           \normsqr[\Vk] {d u} - \lambda_1 \normsqr[\Vk] u
     \Bigr].
  \end{multline*}
  Applying the min-max principle for the first eigenvalue of the
  manifold $F$ and $\Vk$, respectively, we conclude
  \begin{equation}
  \label{eq:low.mm}
     \normsqr[F]{d_y u(x,\cdot)} \ge 
        \lambda_1 \normsqr[F] {u(x,\cdot)} \quad \text{and} \quad
     \normsqr[\Vk] {d u} \ge
        \EWDN 1 \Vk \normsqr[\Vk] u.
  \end{equation}
  Note that $u \restr \Vk$ lies in the quadratic form domain of
  $\laplacianDN \Vk$. Using Assumption~\eqref{eq:small} we see that
  $q_\eps(u) \ge 0$.
\end{proof}

We set
\begin{displaymath}
  \normsqr[\eps,1] u :=
  \normsqr[\eps] u + q_\eps(u) =
  \normsqr[\eps] u + (h_\eps(u) - \frac {\lambda_1}{\eps^2}\normsqr[\eps] u).
\end{displaymath}

Let us now formulate a simple consequence of the min-max
principle~\eqref{eq:min.max} which will be crucial in order to compare
eigenvalues of operators acting in different Hilbert spaces (for a
proof, see e.g.~\cite[Lemma~2.1]{exner-post:pre03}.  Suppose that
$\HS$, $\HS'$ are two separable Hilbert spaces with the norms $\norm
\cdot$ and $\norm \cdot '$. We need to compare eigenvalues $\lambda_k$
and $\lambda'_k$ of self-adjoint operators $Q$ and $Q'$ where $Q \ge -
\Lambda$ for some constant $\Lambda \ge 0$, with purely discrete
spectra defined via quadratic forms $q$ and $q'$ on $\mathcal D
\subset \HS$ and $\mathcal D' \subset \HS'$. We set $\normsqr[1] u :=
(1+ \Lambda) \normsqr u + q(u)$.

\begin{lemma}
  \label{lem:main}
  Suppose that $\map J {\mathcal D}{\mathcal D'}$ is a linear map such
  that there exist constants $\delta_1, \delta_2 \ge 0$ with $\delta_1
  < 1/(1 + \Lambda + \lambda_k)$ and
  \begin{align}
    \label{eq:est.norm}
    \normsqr u & \le {\norm{J u}'}^2 + \delta_1 \normsqr[1] u\\
    \label{eq:est.quad}
    q(u) & \ge \, q'(J u) - \delta_2 \normsqr[1] u
  \end{align}
  for all $u \in \mathcal D$. Then
  \begin{displaymath}
    \lambda_k \ge \lambda_k' - \eta_k
  \end{displaymath}
  where $\eta_k$ is a positive function given by
  \begin{equation}
    \label{eq:eta.k}
    \eta_k =
    \eta(\lambda_k, \delta_1, \delta_2) :=
    \frac {(\lambda_k \delta_1 + \delta_2)(1+\Lambda+\lambda_k)}
                    {1-(1+\Lambda+\lambda_k) \delta_1}.
  \end{equation}
  In particular, $\eta_k \to 0$ as $\delta_1, \delta_2 \to 0$.
\end{lemma}

\section{Convergence of the eigenvalues: small vertex neighbourhoods}
\label{sec:graph}

In this section we consider a graph with straight edges approximated
by an open set $\Meps$ as defined in the previous section (the case of
curved edges will be treated in the next section).  We apply the
abstract comparison result~Lemma~\ref{lem:main} to our concrete
problem in order to show an upper and a lower bound on $\EW k {Q_\eps}
= \EW k {\laplacianD \Meps} - \lambda_1/\eps^2 = \EW k \eps -
\lambda_1/\eps^2$.

\subsubsection*{Upper bound}
\label{sec:upper}
We define the linear map $\map {J_0} {\dom q_0} {\dom q_\eps}$ transmitting
(eigen-)functions on the graph to functions on $\Meps$ by
\begin{equation}
  \label{eq:def.j.upper}
  (J_0 f)(z) := \eps^{-m/2}
  \begin{cases}
    f(x)\phi(y), & z=(x,y) \in \Uj\\
    0,           & z \in \Vk
  \end{cases}
\end{equation}
where $\phi$ is the first normalised Dirichlet eigenfunction on the
transversal direction $F$, i.e.,
\begin{displaymath}
  \laplacianD F \phi = \lambda_1 \phi.
\end{displaymath}
Note that $f \restr {\bd \Ij}$ vanishes and therefore $J_0 f \in \dom
q_\eps = \Sobn \Meps$. We begin with the verification
of~\eqref{eq:est.norm} and~\eqref{eq:est.quad}. We have
\begin{equation}
  \label{eq:est.norm1}
  \normsqr[0] f - \normsqr[\eps] {J_0 f}  =
  \eps \sum_{j \in J} a_j \int_{\Ij} |f(x)|^2 \, dx = 
   O(\eps) \normsqr[0] f
\end{equation}
since $\norm[F] \phi = 1$. Furthermore,
\begin{multline}
  \label{eq:est.quad1}
  q_\eps(J_0f) - q_0(f) \\=
  \sum_{j \in J}
  \Bigl[
      \frac {a_j \eps}{1-a_j \eps}
    \int_{\Ij} |f'|^2 \, dx +
    \frac 1 {\eps^2}
    \int_{\Ij} \int_F
    \bigl(
      |d_y \phi|^2 - \lambda_1 |\phi|^2
    \bigr) dy \, |f|^2 (1-a_j \eps) \,dx
  \Bigr].
\end{multline}
Since $\phi$ is the eigenfunction with eigenvalue $\lambda_1$ the
latter integral vanishes and therefore
\begin{equation}
  \label{eq:eq.quad2}
  q_\eps(J_0f) - q_0(f) =
  \sum_{j \in J}
     \frac {a_j \eps}{1-a_j \eps}
    \normsqr[\Ij] {f'} = O(\eps) q_0(f).
\end{equation}
Applying Lemma~\ref{lem:main} with $\Lambda=0$ we obtain
\begin{equation}
  \label{eq:est.upper}
  \EW k \eps - \frac {\lambda_1}{\eps^2} \le \EW k 0 + O(\eps).
\end{equation}

\subsubsection*{Lower bound}
For the lower bound we have to work a little bit harder. We define
$\map{J_\eps}{\dom q_\eps}{\dom q_0}$ by
\begin{equation}
  \label{eq:def.j.lower}
  (J_\eps u)_j(x) := 
    \eps^{m/2} \bigl( Nu(x) - \rho (x) Nu(x^0) \bigr)
\end{equation}
where
\begin{equation}
\label{eq:def.nu}
  Nu(x) := \iprod {u(x,\cdot)}{\phi} = \int_F u(x,y) \overline \phi(y) \, dy
\end{equation}
is the expectation value of $u(x, \cdot) \in \Lsqr F$ corresponding to
the lowest transversal eigenfunction $\phi$. Here, $x^0$ depends on
$x$ and denotes the left resp.\ right endpoint of $\Ij$ if $x$ is in
the left resp.\ right half of $\Ij$. Furthermore, $\rho$ is a smooth
function with $0 \le \rho(x) \le 1$, $\rho(x)=0$ near the mid point
of $\Ij$ and $\rho(x)=1$ near the boundary of $\Ij$. Abusing the
notation a little bit, $x^0$ also represents an element of $\bd \Ij$.
Since $J_\eps u(x^0) = 0$, we have $J_\eps u \in \dom q_0$.

Again, we begin with the verification of~\eqref{eq:est.norm}.
First, we show the following estimate on higher transversal modes.
\begin{lemma}
  \label{lem:transversal}
  We have
  \begin{displaymath}
    \normsqr v - |\iprod v \phi|^2 \le 
    \frac 1 {\lambda_2-\lambda_1}       
       \bigl( 
           \normsqr {dv} - \lambda_1 \normsqr v
       \bigr)
  \end{displaymath}
  for $v \in \Sobn F$ where $\lambda_i$ are the Dirichlet eigenvalues
  of $F$.
\end{lemma}
\begin{proof}
  Since $v - \iprod v \phi \phi$ is the projection onto $\phi^\orth$,
  the min-max principle implies
  \begin{multline*}
     \normsqr v - |\iprod v \phi|^2 =
     \normsqr {v - \iprod v \phi \phi} \le
     \frac 1 {\lambda_2} \normsqr {d(v - \iprod v \phi \phi)} \\ =
     \frac 1 {\lambda_2}
       \bigl( 
           \normsqr {dv} - \lambda_1 |\iprod v \phi|^2
       \bigr) =
     \frac 1 {\lambda_2}
       \bigl( 
           \normsqr {dv} - \lambda_1 \normsqr v
       \bigr) +
     \frac {\lambda_1} {\lambda_2}
       \bigl( 
           \normsqr v - |\iprod v \phi|^2
       \bigr).
  \end{multline*}
  Since $F$ is connected, $\lambda_1/\lambda_2<1$ and we can bring the
  last difference on the LHS, divide by $(1 - \lambda_1/\lambda_2)$
  and obtain the desired estimate.
\end{proof}

The next lemma shows that under our main assumption, eigenfunctions do
not concentrate at the vertex neighbourhoods:
\begin{lemma}
  \label{lem:no.conc}
  Assume~\eqref{eq:small} then
  \begin{displaymath}
    \normsqr[\Vepsk] u \le
    \frac{\eps^2}{\EWDN 1 \Vk - \lambda_1}
      \Bigl[
         \normsqr[\Vepsk] {du} - \frac{\lambda_1}{\eps^2} \normsqr[\Vepsk] u
      \Bigr]
  \end{displaymath}
  for all $u \in \Sobn \Meps \cap \Sob \Vepsk$.
\end{lemma}
\begin{proof}
  Using the second estimate in~\eqref{eq:low.mm} and the scaling of
  the metric~\eqref{eq:metric} we have
  \begin{multline*}
    \normsqr[\Vepsk] u \le
    \frac{\eps^2}{\EWDN 1 \Vk} \normsqr[\Vepsk] {du} =
    \frac{\eps^2}{\EWDN 1 \Vk}
      \Bigl[
        \normsqr[\Vepsk] {du} - \frac{\lambda_1}{\eps^2} \normsqr[\Vepsk] u
      \Bigr]
      + \frac{\lambda_1}{\EWDN 1 \Vk} \normsqr[\Vepsk] u.
  \end{multline*}
  By our main assumption~\eqref{eq:small}, $\lambda_1/\EWDN 1 \Vk < 1$
  and the result follows as before.
\end{proof}

Finally, we need the following lemma.
\begin{lemma}
  \label{lem:trace}
  We have
  \begin{displaymath}
    \eps^m |Nu(x^0)|^2 \le
    O(\eps)
      \bigl[
        \normsqr[\Vepsk] {du} - \frac{\lambda_1}{\eps^2} \normsqr[\Vepsk] u
      \bigr]
  \end{displaymath}
  for all $u \in \Sobn \Meps \cap \Sob \Vepsk$ and $x^0 \in \bd \Ij$,
  if $e_j$ emanates from $v_k$.
\end{lemma}
\begin{proof}
  A standard Sobolev embedding theorem gives
  \begin{displaymath}
    |Nu(x^0)|^2 \le
    \int_F |u(x^0,y)|^2 dy \le
    c_1
      \bigl[
        \normsqr[\Vk] {du} + \normsqr[\Vk] u
      \bigr]
  \end{displaymath}
  for some constant $c_1>0$ (note that $F=\bd_j \Vk$). Now by the
  scaling of the metric on $\Vk$
  \begin{displaymath}
    \normsqr[\Vk] {du} + \normsqr[\Vk] u =
    \eps^{-m}
      \Bigl[
        \eps
        \bigl(
        \normsqr[\Vepsk] {du} - \frac{\lambda_1}{\eps^2} \normsqr[\Vepsk] u
        \bigr)
        + \frac 1 \eps (1+\lambda_1) \normsqr[\Vepsk] u 
      \Bigr]
  \end{displaymath}
  and the result follows from the preceeding lemma.
\end{proof}

Now, we want to consider the norm difference
\begin{multline*}
  \normsqr[\eps] u - \normsqr[0] {J_\eps u} \\=
  \sum_{k \in K} \normsqr[\Vepsk] u +
  \sum_{j \in J}
    \Bigl[
      \normsqr[\Uepsj] u - \int_{\Ij} |Nu(x)-\rho(x) Nu(x^0)|^2 \eps^m dx
    \Bigr].
\end{multline*}
The first sum can be estimated by $O(\eps^2) q_\eps(u)$ using
Lemma~\ref{lem:no.conc}. For the second, we use
\begin{equation}
    \label{eq:quad.lower}
    (a+b)^2 \ge (1-\delta)a^2 - \frac 1 \delta b^2, \qquad \delta>0
\end{equation}
and obtain as upper bound
\begin{multline*}
  \normsqr[\Uepsj] u - (1-\delta)\int_{\Ij} |Nu(x)|^2 \eps^m dx +
    \frac {\eps^m} \delta \int_{\Ij} |\rho(x)|^2 dx \, 
                \max_{x^0 \in \bd \Ij}|Nu(x^0)|^2 \\
      \le
  \int_{\Ij}
   \Bigl[
      \normsqr{u(x,\cdot)} - |\iprod {u(x,\cdot)} \phi|^2
   \Bigr] \eps^m dx \\ +
   \Bigl( \frac \delta {1-a_j\eps} - a_j \eps \Bigr)
      \normsqr[\Uepsj] u
   + \frac {\eps^m} \delta \normsqr[\Ij] \rho \, 
                \max_{x^0 \in \bd \Ij} |Nu(x^0)|^2
\end{multline*}
using Cauchy-Schwarz. Applying Lemma~\ref{lem:transversal}, the
scaling of the metric on $F$ in~\eqref{eq:metric},
Lemma~\ref{lem:trace} and setting $\delta=\eps^{1/2}$, we end up with
the estimate
\begin{equation}
  \label{eq:est.norm3}
  \normsqr[\eps] u - \normsqr[0] {J_\eps u} \le
  O(\eps^{1/2}) \normsqr[\eps,1] u.
\end{equation}

For the quadratic form difference we have
\begin{multline*}
  q_0(J_\eps u) - q_\eps(u) \\ \le
  \sum_{j \in J}
    \eps^m \normsqr[\Ij] {N (\partial_x u) - \rho' Nu(x^0)} -
    \frac 1{1-a_j \eps} \int_{\Ij} \normsqr{\partial_x u(x,\cdot)} \eps^m dx
\end{multline*}
where the terms of order $\eps^{-2}$ has been estimated
with~\eqref{eq:low.mm}. Using
\begin{equation}
  \label{eq:quad.upper}
  (a+b)^2 \le (1+\delta)a^2 + \frac 2 \delta b^2, \qquad 0 < \delta \le 1,
\end{equation}
with $\delta=\eps^{1/2}$, Cauchy-Schwarz for $|N (\partial_x
u(x,\cdot)|^2 \le \normsqr{\partial_x u(x,\cdot)}$ and
Lemma~\ref{lem:trace}, we obtain
\begin{equation}
  \label{eq:est.quad3}
  q_0(J_\eps u) - q_\eps(u) \\ \le
  O(\eps^{1/2}) \normsqr[\eps,1] u.  
\end{equation}
Applying Lemma~\ref{lem:main} again (with $\Lambda=0$), we obtain
\begin{equation}
  \label{eq:est.lower}
  \EW k {Q_\eps} = \EW k \eps - \frac {\lambda_1}{\eps^2} \ge 
  \EW k 0 - \eta_k.
\end{equation}
Here, $\eta_k = O(\eps^{1/2})$
using~\eqref{eq:est.norm3},~\eqref{eq:est.quad3} and the upper
estimate $\EW k {Q_\eps} \le \EW k 0 + O(\eps) = O(1)$
from ~\eqref{eq:est.upper}.

\section{Curved edges}
\label{sec:curvature}

Let us now consider a \emph{curved} quantum wave guide embedded in
$\R^2$ (more general embeddings can be treated similarly). Such spaces
have already been analysed e.g.\ in~\cite{exner-seba:89}
or~\cite{duclos-exner:95}. We only consider a single edge here since
one can easily replace the edge estimates without curvature by the
appropriate estimates with curvature in the previous
section\footnote{More precisely: one has to show the estimates of this
  section for the metric
  \begin{displaymath}
    g_\eps = (1- \eps a)^2(1+ \eps y \, \kappa(x))^2 dx^2 + \eps^2 dy^2    
  \end{displaymath}
  instead of the metric defined in~\eqref{eq:met.curv} in order to
  take into account the shortened edges due to the embedding. To keep
  the notation manageable we omit this fact here.} (cf.\ 
Remark~\ref{rem:graph.curv} for the precise assumptions on the
curvature).  The convergence of the discrete spectrum of an infinite
curved quantum wave guide has already been established in
\cite{duclos-exner:95} using perturbation arguments and an asymptotic
expansion (cf.~also \cite{karp-pinsky:88} where the asymptotic of the
first Dirichlet eigenvalue of a $\eps$-neighbourhood of a finite
length curve in $\R^3$ was treated).  Here, in contrast, we use the
variational arguments of Lemma~\ref{lem:main} which are somehow
simpler (the price being a weaker result).

\subsubsection*{Definition of the approximating space}
Suppose that $\map \gamma I {\R^2}$ is a smooth curve (e.g. $C^4$ is
enough) with bounded derivatives parametrised by arclength (i.e. the
tangent vector $\dot \gamma(x)$ has unit length for all $x \in I$).
Suppose that either $\gamma$ is a closed curve ($I \cong \Sphere^1$)
or has two ends ($I \cong (0,1)$).

We introduce the $\eps$-neighbourhood $\Ueps$ of the curve given as
the image of the parametrisation
\begin{equation}
  \label{eq:param}
  \begin{array}{rl}
    \map \Psi {I \times F & } {\Ueps \subset \R^2}\\
    (x,y) & \longmapsto \gamma(x) + \eps  y \, n(x)
  \end{array}
\end{equation}
where $n(x):=(\dot \gamma_2(x),-\dot \gamma_1(x))$ is orthogonal to
the tangent vector $\dot \gamma(x)$ and $F=(-1,1)$. Define the signed
curvature by
\begin{equation}
  \label{eq:curv}
  \kappa := 
    \dot \gamma_1 \ddot \gamma_2 - \dot \gamma_2 \ddot \gamma_1  
\end{equation}
and suppose that $0 < \eps \le \eps_0 := 1/(2\norm[\infty] \kappa)$
where $\norm[\infty] \kappa$ denotes the supremum of $|\kappa(x)|$, $x
\in I$. We assume in addition that $\Psi$ is a diffeomorphism.
\begin{remark}
  \label{rem:graph.curv}
  Suppose we consider an embedded graph $M_0$ with \emph{curved} edges
  $e_j$ with curvature $\kappa_j$. Besides the assumption that the
  parametrisation~\eqref{eq:param} is a diffeomorphism for each edge
  $e_j$ we need the additional hypothesis that $\supp \kappa_j$ is
  contained in the open interval $I_j$, i.e., that the curvature
  vanishes in a small neighbourhood of the adjacent vertices.
  Otherwise one needs to modify the scaling property of the vertex
  neighbourhoods $\Vepsk$: they cannot be homothetic to a fixed set
  $\Vk$ if the edge is curved up to the vertex $v_k$.
\end{remark}

Denote by $g_\eps$ the pull-back of the Euclidean metric via $\Psi$,
i.e., $g_\eps := \Psi^* g_{\mathrm{eucl}}$. A straightforward
calculation shows that
\begin{equation}
  \label{eq:met.curv}
  g_\eps = (1+ \eps y \, \kappa(x))^2 dx^2 + \eps^2 dy^2.
\end{equation}
We denote by $\Ueps$ the manifold $I \times F$ with metric $g_\eps$ and
by $\tUeps$ the same manifold with the product metric
\begin{displaymath}
  \widetilde g_\eps = dx^2 + \eps^2 dy^2. 
\end{displaymath}
For computational reasons, it is much
easier to deal with the latter metric so we introduce the unitary
transformation
\begin{equation}
  \label{eq:un.trafo}
  \begin{array}{rl}
    \map \Phi {\Lsqr \Ueps &} {\Lsqr \tUeps}\\
    u & \longmapsto (1+ \eps y \, \kappa(x))^{1/2} u.
   \end{array}
\end{equation}
Note that $\det g_\eps^{1/2} = \eps(1+\eps y \, \kappa(x)) > 0$ is the
density of the metric $g_\eps$ and $\det \widetilde g_\eps^{1/2} =
\eps$.  For the rest of the section, we will work in the Hilbert space
$\HS_\eps := \Lsqr \tUeps$.
\subsubsection*{Definition of the operator on the thickened set}
We want to consider the Dirichlet Laplacian on $\Ueps$. Its quadratic
form is $\normsqr[\Ueps]{du}$, $u \in \Sobn \Ueps$ (we could also
allow other boundary conditions at $\bd I \times F$). The transformed
quadratic form is given by
\begin{displaymath}
  h_\eps(u):=
  \normsqr[\Ueps]{d\Phi^*u} =
  \normsqr[\Ueps]{d((1+ \eps y \, \kappa)^{-1/2} u)}, \qquad
  u \in \Sobn \tUeps = \Sobn \Ueps
\end{displaymath}
A straightforward calculation already performed at various places
(e.~g.~\cite{exner-seba:89} or~\cite{duclos-exner:95})
yields
\begin{equation}
  \label{eq:quad.trafo}
  h_\eps(u) = 
  \int_I \int_F 
    \Bigl[
      \frac 1 {(1+ \eps y \, \kappa)^2} |\partial_x u|^2 +
      \frac 1 {\eps^2} |\partial_y u|^2
      + K_\eps |u|^2
    \Bigr]
    \eps \, dy\, dx
\end{equation}
where the curvature induced potential $K_\eps$ is given by
\begin{equation}
  \label{eq:curv.pot}
  K_\eps(x,y) = 
   -\frac {\kappa^2} {4(1+\eps y \, \kappa)^2}
    + \frac {\eps y \, \ddot \kappa} {2(1+\eps y \, \kappa)^3}
    - \frac {5 \eps^2 y^2 \, \dot \kappa^2} {4(1+\eps y \, \kappa)^4}.
\end{equation}
Note that $K_\eps$, $0 < \eps \le \eps_0$, is bounded from below by a
constant $-\Lambda_{\eps_0}$, $\Lambda_{\eps_0} \ge 0$ depending only
on the supremum of $\kappa$, $\dot \kappa$, $\ddot \kappa$ and
$\eps_0$. Using the first estimate in~\eqref{eq:low.mm} we see that
\begin{equation}
  \label{eq:quad.scaled}
  q_\eps(u) := h_\eps(u) - \frac {\lambda_1}{\eps^2} \normsqr[\tUeps] u
\end{equation}
is bounded from below by $-\Lambda_{\eps_0} \normsqr[\tUeps] u$.
Therefore,
\begin{displaymath}
  \normsqr[\eps,1] u := q_\eps(u) + (\Lambda_{\eps_0}+1) \normsqr[\tUeps] u 
\end{displaymath}
defines a norm on the quadratic form domain $\Sobn \tUeps$.

\subsubsection*{Definition of the limit space and operator}
Finally, we define the limit operator $Q_0$. Clearly, $Q_0$ will act
in the limit space $\HS_0:=\Lsqr I$. As usual, we define $Q_0$ via its
quadratic form
\begin{equation}
  \label{eq:quad.lim}
  q_0(f):=
  \int_I \Bigl[ |f'|^2 - \frac {\kappa^2} 4 |f|^2 \Bigr] \, dx.
\end{equation}
Again,
\begin{displaymath}
  \normsqr[0,1] f := 
   q_0(f) + \Bigl(\frac{\norm[\infty] \kappa^2} 4 + 1 \Bigr) \normsqr[I] f 
\end{displaymath}
defines a norm on the quadratic form domain $\Sobn I$.  Note that
$K_\eps(x,y) = - \kappa(x)^2/4 + O(\eps)$ as $\eps \to 0$.

\subsubsection*{Spectral convergence}
We want to show the following spectral convergence. From its proof it
is straightforward to show Theorem~\ref{thm:main} in the general case
of branched quantum wave guides with curved edges.
\begin{theorem}
  \label{thm:curved.edge}
  Denote by $\EW k \eps$ the $k$-th Dirichlet eigenvalue of the curved quantum
  wave guide $\Ueps$. Then
  \begin{displaymath}
    \EW k \eps - \frac{\lambda_1}{\eps^2} = \EW k 0 + O(\eps), \qquad
    \eps \to 0,
  \end{displaymath}
  where $\EW k 0$ denotes the $k$-th eigenvalue of $Q_0 = -d^2/d^2x -
  \kappa^2/4$.
\end{theorem}
Here, $\lambda_1=\pi^2/4$ is the first Dirichlet eigenvalue of
$F=(-1,1)$.  As before, we establish an upper and a lower bound on
$\EW k \eps$.

\subsubsection*{Upper bound}
We define the transition operator $J_0$ as in \eqref{eq:def.j.upper}
on the edges (here, $m=1$). Clearly, we have
\begin{displaymath}
  \normsqr[0] f = \normsqr[\eps] {J_0 f}
\end{displaymath}
since $\phi$ is supposed to be normalised. In addition,
\begin{multline*}
  q_\eps(J_0 f) - q_0(f) = \\
  \int_I \int_F 
    \Bigl[
      \bigl( \frac 1 {(1+ \eps y \, \kappa)^2} - 1 \bigr) 
      |f'\phi|^2 +
      |f|^2 \frac 1 {\eps^2} \bigl( |\phi'|^2 - \lambda_1 |\phi|^2 \bigr)
      + \bigl( K_\eps + \frac {\kappa^2} 4 \bigr)|f \phi|^2
    \Bigr]
    \eps \, dy\, dx
\end{multline*}
which can be estimated by $O(\eps) \normsqr[0,1] f$ where $O(\eps)$
depends only on $\kappa$ (and its derivatives) (remember that $\phi$
is the first Dirichlet eigenfunction on $F$ with eigenvalue
$\lambda_1$). Applying Lemma~\ref{lem:main} yields the desired upper
estimate with $\eta_k(\eps) = O(\eps)$.

\subsubsection*{Lower bound}
The lower bound is again a little bit more difficult. We define the
transition operator $J_\eps$ by
\begin{equation}
  \label{eq:j.lower.curv}
  (J_\eps u)(x) := 
    \eps^{1/2} Nu(x)
\end{equation}
where $Nu$ is the transversal expectation value of $u$ with respect to
$\phi$, cf.\ Equation~\eqref{eq:def.nu}.  Applying
Lemma~\ref{lem:transversal} for $v=u(x,\cdot)$ we obtain the estimate
\begin{multline*}
  \normsqr[\eps] u - \normsqr[0] {J_\eps u} \le
  \frac{\eps^2}{\lambda_2 - \lambda_1}
  \int_I \int_F
    \Bigl[
      \frac 1 {\eps^2}
      \bigl( |\partial_y u|^2 - \lambda_1 |u|^2 \bigr)
    \Bigr]
    \eps \, dy\, dx \\ \le
  \frac{\eps^2}{\lambda_2 - \lambda_1}
    \bigl( q_\eps(u) + \Lambda_{\eps_0} \normsqr[\eps] u \bigr) =
  O(\eps^2) \normsqr[\eps,1] u.
\end{multline*}
The quadratic form estimate is given by
\begin{multline*}
  q_0(J_\eps u) - q_\eps(u) =
  \int_I \int_F
    \Bigl[
      \bigl(
         |\iprod{\partial_x u(x, \cdot)} \phi |^2 -
           \frac 1 {(1+\eps y \, \kappa)^2} |\partial_x u|^2
      \bigr) - {} \\
      \frac 1{\eps^2}
      \bigl(
         | \partial_y u |^2 - \lambda_1 |u|^2
      \bigr) 
      + \frac {\kappa^2} 4
      \bigl(
          |u|^2 - |\iprod{u(x, \cdot)} \phi|^2 
      \bigr) 
      - |u|^2
      \Bigl(
         \frac {\kappa^2} 4 + K_\eps 
      \Bigr)
    \Bigr]
    \eps \, dy\, dx.
\end{multline*}
As before, we estimate the first difference using Cauchy-Schwarz. The
second difference is negative (cf.~\eqref{eq:low.mm}). The third
difference is small due to Lemma~\ref{lem:transversal}. The forth
difference is also small since $K_\eps = -\kappa^2/4 + O(\eps)$.
Therefore, we end up with an upper estimate given by $O(\eps)
\normsqr[\eps,1] u$.  Applying Lemma~\ref{lem:main} once more, we
obtain the desired lower estimate on $\EW k \eps$.  Using also the
upper estimate we see that $\eta_k(\eps)=O(\eps)$.

\section{Examples}
\label{sec:examples}

In this section, we want to comment on the smallness
condition~\eqref{eq:small} and give examples where this condition
holds or fails. To simplify the presentation, we assume that $M_0
\subset \R^2$.

\begin{figure}[h]
  \begin{center}
    \includegraphics{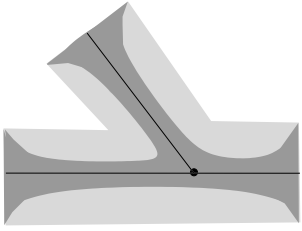}
    \caption{The original vertex neighbourhood $\Vk(0)$ (light grey) and 
     the shrunken vertex neighbourhood $\Vk(\tau)$ (dark grey).}
    \label{fig:vertex2}
  \end{center}
\end{figure}

First, we show, that the condition can always be fulfilled, provided
the vertex neighbourhood is small enough. Suppose that we start with
the $1$-neighbourhood denoted by $\Vk(0)$, i.e., we set $\eps=1$ and
regard the unscaled vertex neighbourhood $\Vk$. Remember that we have
assumed that the curvature vanishes near the vertices, therefore
$\Vk(0)$ is bounded by straight lines. Then we deform $\Vk(0)$
smoothly in order to obtain a family $\Vk(\tau)$, $\tau \ge 0$,
shrinking to the graph, but fixing the boundary parts $\bd_j
\Vk(\tau)=\bd_j \Vk(0)$, $j \in J_k$, where the edge neighbourhoods
touch (cf.~Figure~\ref{fig:vertex2}).
As in~\cite[Sec.~7]{post:03b} we can show that the first eigenvalue of
the Laplacian on $\Vk(\tau)$ with Dirichlet boundary conditions except
on the fixed boundary parts $\bd_j \Vk(\tau)$, where we impose Neumann
boundary conditions, tends to $\infty$, i.e.,
\begin{displaymath}
  \EWDN 1 {\Vk(\tau)} \to \infty \qquad \text{as $\tau \to \infty$.} 
\end{displaymath}
Therefore there always exists a fixed $\tau \in (0,\infty)$ such that
$\Vk:=\Vk(\tau)$ satisfies~\eqref{eq:small}. Fixing this shrinking
parameter $\tau$, we proceed with the definition of $\Meps$ as in
Section~\ref{sec:prelim}.

\subsubsection*{An example not satisfying the smallness assumption}
Let us briefly give an example of a vertex neighbourhood not
satisfying~\eqref{eq:small}. For suitable vertex neighbourhoods (e.g.\
arising from the $\eps$-neighbourhood of a graph) we will show the
existence of an eigenvalue below the threshold $\lambda_1/\eps^2 =
\pi^2/(4\eps^2)$ (cf.\ also ~\cite{srw:89} and~\cite{abgm:91} for the
case of an $\eps$-neighbourhood of a vertex with four infinite edges
emanating (a ``cross''); in the former reference one can also find a
contour plot of the first eigenfunction). Therefore, the conclusion of
Theorem~\ref{thm:main} is false, i.e.,~\eqref{eq:small} fails.

\begin{figure}[h]
  \begin{center}
\begin{picture}(0,0)%
\includegraphics{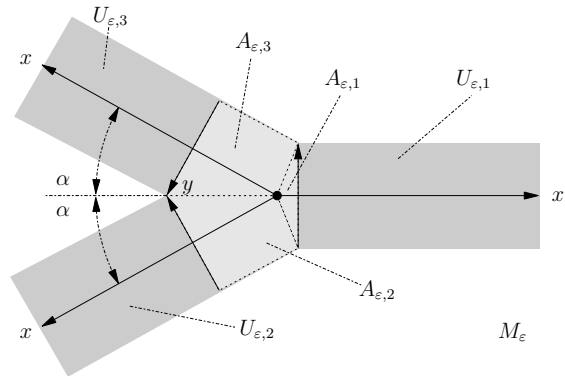}%
\end{picture}%
\setlength{\unitlength}{4144sp}%
\begin{picture}(4866,3167)(406,-2511)
\put(3106,-16){$A_{\eps,1}$}
\put(4231,-16){$U_{\eps,1}$}
\put(2380,-2150){$U_{\eps,2}$}
\put(3400,-1816){$A_{\eps,2}$}
\put(2341,299){$A_{\eps,3}$}
\put(1126,524){$U_{\eps,3}$}
\put(500,164){$x$}
\put(500,-2176){$x$}
\put(5050,-1006){$x$}
\put(1891,-916){$y$}
\put(811,-1141){$\alpha$}
\put(811,-871){$\alpha$}
\put(4591,-2176){$\Meps$}
\end{picture}%
    \caption{
      A simple trial function supported in a neighbourhood of the
      vertex has an eigenvalue below the threshold
      $\lambda_1/\eps^2=\pi^2/(4\eps^2)$.}
    \label{fig:counterex}
  \end{center}
\end{figure}

The existence of such an eigenvalue below the threshold can easily be
established by inserting an appropriate trial function in the Rayleigh
quotient. We consider a graph with one vertex and three adjacent edges
of length $\ell$ and denote its $\eps$-neighbourhood by $\Meps$. We
decompose $\Meps$ into three rectangles $\Uepsj$ and three sets
$\Aepsj$ as in Figure~\ref{fig:counterex}.

On the rectangle $\Uepsj$ we use the coordinates $0 < x < \ell$ and
$-\eps < y < \eps$ where $x=0$ corresponds to the common boundary with
$\Aepsj$. We extend these coordinates from $\Uepsj$ onto $\Aepsj$ and
define
\begin{displaymath}
  u(x,y):= \eps^{-1/2}\chi(x)\cos (\frac \pi {2\eps} y)
\end{displaymath}
as a test function on each of the three sets $\Uepsj \cup \Aepsj$.
Here, $\chi(x)=1$ for $x<0$ (i.e., on $\Aepsj$), $\chi(x)=\cos(\pi
x/(2 \eps \kappa))$ for $0 \le x \le \kappa \eps$ and $\chi(x)=0$ for
$\eps < x < \ell$ where $\kappa>0$ is some constant to be specified
later.  Although $u$ is not differentiable across the different
borders (but continuous), it still lies in the quadratic form domain
$\Sobn \Meps$.

A straightforward calculation yields
\begin{equation}
\label{eq:angle}
  \frac {\normsqr[\Meps] {du}} {\normsqr[\Meps] u} - \frac {\pi^2} {4\eps^2} =
  \Big(
     \frac {8\kappa \cos \alpha + 3\pi^2 \sin \alpha - 16\kappa}
           {((3\pi^2-4)\cos \alpha + 3\pi^2 \kappa \sin \alpha + 8)\kappa} 
  \Big)
  \frac {\pi^2} {4\eps^2}. 
\end{equation}
This quantity is negative for all $0<\alpha < 0.93\pi$ if we choose
e.g.\ $\kappa=3$. In particular, there exists a negative eigenvalue of
$\laplacianD \Meps - \lambda_1/\eps^2$ of order $\eps^{-2}$, and
Condition~\eqref{eq:small} fails here for any choice of vertex
neighbourhoods $\Vepsk$ since the conclusion of Theorem~\ref{thm:main}
is false.  Note that the vertex neighbourhoods $\Vepsk$ are not
uniquely determined.  One could enlarge $\Vepsk$ at each edge
emanating by a cylinder of length $a \eps$ taken away from the
corresponding edge neighbourhood.

If $\laplacianD \Meps - \lambda_1/\eps^2$ has negative eigenvalues it
is not clear whether its appropriately scaled eigenvalues converge to
eigenvalues of an operator on the graph $M_0$. The dependence of the
leading order on the angle $\alpha$ in~\eqref{eq:angle} indicates that
the limit should depend on the angles of the edges meeting at a
vertex.

\section*{Acknowledgements}

It is a pleasure to thank Pierre Duclos for the invitation at the
Centre de Physique Th\'eorique (Marseille-Luminy) where this work has
been initiated. In addition, the author would appreciate Pavel Exner
and Daniel Grieser for fruitful discussions on this problem.



\def\cprime{$'$} \def\cprime{$'$} \def\cprime{$'$}
\providecommand{\bysame}{\leavevmode\hbox to3em{\hrulefill}\thinspace}
\providecommand{\MR}{\relax\ifhmode\unskip\space\fi MR }
\providecommand{\MRhref}[2]{%
  \href{http://www.ams.org/mathscinet-getitem?mr=#1}{#2}
}
\providecommand{\href}[2]{#2}
\begin{thebibliography}{ABGM91}

\bibitem[ABGM91]{abgm:91}
Y.~Avishai, D.~Bessis, B.~G. Giraud, and G.~Mantica, \emph{Quantum bound states
  in open geometries}, Phys. Rev. B \textbf{44} (1991), no.~15, 8028--8034.

\bibitem[Dav96]{davies:96}
E.~B. Davies, \emph{{Spectral theory and differential operators}}, Cambridge
  University Press, Cambridge, 1996.

\bibitem[DE95]{duclos-exner:95}
P.~Duclos and P.~Exner, \emph{Curvature-induced bound states in quantum
  waveguides in two and three dimensions}, Rev. Math. Phys. \textbf{7} (1995),
  no.~1, 73--102. \MR{MR1310767 (95m:81239)}

\bibitem[EP04]{exner-post:pre03}
P.~Exner and O.~Post, \emph{{Convergence of spectra of graph-like thin
  manifolds}}, to appear in Journal of Geometry and Physics (2004).

\bibitem[E{\v{S}}89]{exner-seba:89}
P.~Exner and P.~{\v{S}}eba, \emph{Bound states in curved quantum waveguides},
  J. Math. Phys. \textbf{30} (1989), no.~11, 2574--2580.

\bibitem[Kat66]{kato:66}
T.~Kato, \emph{Perturbation theory for linear operators}, Springer-Verlag,
  Berlin, 1966.

\bibitem[KP88]{karp-pinsky:88}
L.~Karp and M.~Pinsky, \emph{First-order asymptotics of the principal
  eigenvalue of tubular neighborhoods}, Geometry of random motion (Ithaca,
  N.Y., 1987), Contemp. Math., vol.~73, Amer. Math. Soc., Providence, RI, 1988,
  pp.~105--119. \MR{MR954634 (89g:58206)}

\bibitem[KS99]{kostrykin-schrader:99}
V.~Kostrykin and R.~Schrader, \emph{Kirchhoff's rule for quantum wires}, J.
  Phys. A \textbf{32} (1999), no.~4, 595--630.

\bibitem[Kuc02]{kuchment:02}
P.~Kuchment, \emph{Graph models for waves in thin structures}, Waves Random
  Media \textbf{12} (2002), no.~4, R1--R24.

\bibitem[Kuc04]{kuchment:04}
\bysame, \emph{Quantum graphs: {I}. {S}ome basic structures}, Waves Random
  Media \textbf{14} (2004), S107--S128.

\bibitem[KZ01]{kuchment-zeng:01}
P.~Kuchment and H.~Zeng, \emph{Convergence of spectra of mesoscopic systems
  collapsing onto a graph}, J. Math. Anal. Appl. \textbf{258} (2001), no.~2,
  671--700.

\bibitem[KZ03]{kuchment-zeng:03}
\bysame, \emph{Asymptotics of spectra of {N}eumann {L}aplacians in thin
  domains}, Advances in differential equations and mathematical physics
  (Birmingham, AL, 2002), Contemp. Math., vol. 327, Amer. Math. Soc.,
  Providence, RI, 2003, pp.~199--213. \MR{1 991 542}

\bibitem[Pos03]{post:03b}
O.~Post, \emph{Eigenvalues in spectral gaps of a perturbed periodic manifold},
  Mathematische Nachrichten \textbf{261--262} (2003), 141--162.

\bibitem[RS53]{ruedenberg-scherr:53}
K.~Ruedenberg and C.~W. Scherr, \emph{{Free--electron network model for
  conjugated systems, I.~Theory}}, J. Chem. Phys. \textbf{21} (1953),
  1565--1581.

\bibitem[RS80]{reed-simon-1}
M.~Reed and B.~Simon, \emph{{Methods of modern mathematical physics I:
  Functional analysis}}, Academic Press, New York, 1980.

\bibitem[RS01]{rubinstein-schatzman:01}
J.~Rubinstein and M.~Schatzman, \emph{Variational problems on multiply
  connected thin strips. {I}. {B}asic estimates and convergence of the
  {L}aplacian spectrum}, Arch. Ration. Mech. Anal. \textbf{160} (2001), no.~4,
  271--308. \MR{1 869 667}

\bibitem[SRW89]{srw:89}
R.~L. Schult, D.~G. Ravenhall, and H.~W. Wyld, \emph{Quantum bound states in a
  classically unbound system of crossed wires}, Phys. Rev. B \textbf{39}
  (1989), no.~8, 5476--5479.

\end{thebibliography}
\providecommand{\bysame}{\leavevmode\hbox to3em{\hrulefill}\thinspace}

\end{document}